\documentclass[acus]{JAC2003}
\usepackage{graphicx}
\usepackage{booktabs}

\usepackage{mathptmx}
\usepackage{dblfloatfix}

\newcommand{\q}[2]{\ensuremath{#1\ \mathrm{#2}}}
\newcommand{\code}[1]{#1}

\newcommand{\qp}{\ensuremath{Q_p}}
\newcommand{\qa}{\ensuremath{Q_a}}

\newcommand{\xip}{\ensuremath{\xi_p}}
\newcommand{\xia}{\ensuremath{\xi_a}}

\begin{document}

\title{DETECTION OF COHERENT BEAM-BEAM MODES\\ WITH DIGITIZED BEAM
  POSITION MONITOR SIGNALS}

\author{G.~Stancari\thanks{E-mail: stancari@fnal.gov.} and
  A.~Valishev,
  Fermi National Accelerator Laboratory, Batavia, IL 60510, USA\\
  S.~M.~White, Brookhaven National Laboratory, Upton, NY 11973, USA}

\maketitle

\begin{abstract}
  A system for bunch-by-bunch detection of transverse proton and
  antiproton coherent oscillations in the Fermilab Tevatron collider
  is described. It is based on the signal from a single beam-position
  monitor located in a region of the ring with large amplitude
  functions. The signal is digitized over a large number of turns and
  Fourier-analyzed offline with a dedicated algorithm. To enhance the
  signal, band-limited noise is applied to the beam for about
  1~s. This excitation does not adversely affect the circulating beams
  even at high luminosities. The device has a response time of a few
  seconds, a frequency resolution of $1.6\times 10^{-5}$ in fractional
  tune, and it is sensitive to oscillation amplitudes of 60~nm. It
  complements Schottky detectors as a diagnostic tool for tunes, tune
  spreads, and beam-beam effects. Measurements of coherent mode
  spectra are presented and compared with models of beam-beam
  oscillations.
\end{abstract}

\section{INTRODUCTION}

In particle colliders, each beam experiences nonlinear forces when
colliding with the opposing beam. A manifestation of these forces is a
vibration of the bunch centroids around the closed orbit.  These
coherent beam-beam oscillation modes were observed in several lepton
machines, including PETRA, TRISTAN, LEP, and
VEPP-2M~\cite{Piwinski:IEEE:1979, Ieiri:NIM:1988, Keil:ICHEA:1992,
  Nesterenko:PRE:2002}. Although their observation in hadron machines
is made more challenging by the lack of strong damping mechanisms to
counter external excitations, they were seen at the ISR, at RHIC, in
the Tevatron, and in the LHC~\cite{Koutchouk:CERN:1982a,
  Koutchouk:CERN:1982b, Fischer:BNL:2002, Fischer:PAC:2003,
  Pieloni:PhD:2008, Stancari:PRSTAB:2012, Buffat:IPAC:2011}.
Originally, one motivation for the study of coherent beam-beam modes
was the realization that their frequencies may lie outside the
incoherent tune distribution, with a consequent loss of Landau
damping~\cite{Alexahin:PA:1999}. The goal of the present research is
to develop a diagnostic tool to estimate bunch-by-bunch tune
distributions, to assess the effects of Gaussian electron lenses for
beam-beam compensation~\cite{Shiltsev:PRSTAB:1999,Shiltsev:NJP:2008,
  Shiltsev:PRSTAB:2008, Valishev:PAC:2011}, and to provide an
experimental basis for the development of beam-beam numerical codes.

The behavior of colliding bunches is analogous to that of a system of
oscillators coupled by the beam-beam force. In the simplest case, when
2~identical bunches collide head-on in one interaction region,
2~normal modes appear: a $\sigma$~mode (or 0~mode) at the lattice
tune, in which bunches oscillate transversely in phase, and a
$\pi$~mode, separated from the $\sigma$~mode by a shift of the order
of the beam-beam parameter, in which bunches are out of phase. In
general, the number, frequency, and amplitude of these modes depend on
the number of bunches, on the collision pattern, on the tune
separation between the two beams, on transverse beam sizes, and on
relative intensities. Coherent beam-beam modes have been studied at
several levels of refinement, from analytical linear models to fully
3-dimensional particle-in-cell calculations~\cite{Piwinski:IEEE:1979,
  Pieloni:PhD:2008, Meller:IEEE:1981, Yokoya:PA:1990,
  Herr:PRSTAB:2001, Alexahin:NIM:2002, Pieloni:PAC:2005,
  Qiang:PRSTAB:2002, Qiang:NIM:2006, Stern:PRSTAB:2010}.

In the Tevatron, 36 proton bunches (identified as P1--P36) collided
with 36 antiproton bunches (A1--A36) at the center-of-momentum energy
of 1.96~TeV. There were 2 head-on interaction points (IPs),
corresponding to the CDF and the DZero experiments. Each particle
species was arranged in 3~trains of 12~bunches each, circulating at a
revolution frequency of 47.7~kHz. The bunch spacing within a train was
396~ns, or 21 53-MHz rf buckets. The bunch trains were separated by
2.6-$\mu$s abort gaps. The synchrotron frequency was 34~Hz, or
$7\times 10^{-4}$ times the revolution frequency. The machine operated
with betatron tunes near~20.58.

The betatron tunes and tune spreads of individual bunches are among
the main factors that determine beam lifetimes and collider
performance. They are affected by head-on and long-range beam-beam
interactions. Three systems were used in the Tevatron to measure
incoherent tune distributions: the 21.4-MHz Schottky detectors, the
1.7-GHz Schottky detectors, and the direct diode detection base band
tune (3D-BBQ). The latter two could be gated on single
bunches. Detection of transverse coherent modes complemented these
three systems because of its sensitivity, bunch-by-bunch capability,
high frequency resolution, and fast measurement time.

The basis for the measurement technique was presented in
Ref.~\cite{Carneiro:Beamsdoc:2005}, and preliminary results can be
found in
Refs.~\cite{Semenov:PAC:2007,Kamerdzhiev:BIW:2008,Valishev:EPAC:2008}.
Several improvements, mainly in the data analysis, were implemented
and presented in a concise report~\cite{Stancari:BIW:2010}. A
comprehensive description of the technique and of observations in a
wide range of experimental conditions was reported in
Ref.~\cite{Stancari:PRSTAB:2012}. Here, we focus on the detection of
coherent beam-beam oscillations and on comparisons with analytical and
numerical models.

\section{MODELS}
\label{sec:model}

In the Tevatron, transverse coherent oscillations were substantially
nonlinear due to the properties of the lattice and of the beam-beam
force. We first used the rigid-bunch approximation for a fast analysis
of the expected beam-beam mode frequencies and their dependence on the
the betatron tunes~$Q$ and on the beam-beam parameter per interaction
point~$\xi$. For a more accurate description of the coherent mode
spectrum, tracking simulations with a strong-strong 3-dimensional
numerical code were employed.

\begin{figure}[b!]
\centering
\begin{tabular}[c]{cc}
  \parbox{0.04\columnwidth}{\rotatebox{90}{Fractional tune}} &
  \parbox{0.94\columnwidth}{\includegraphics[width=75mm]{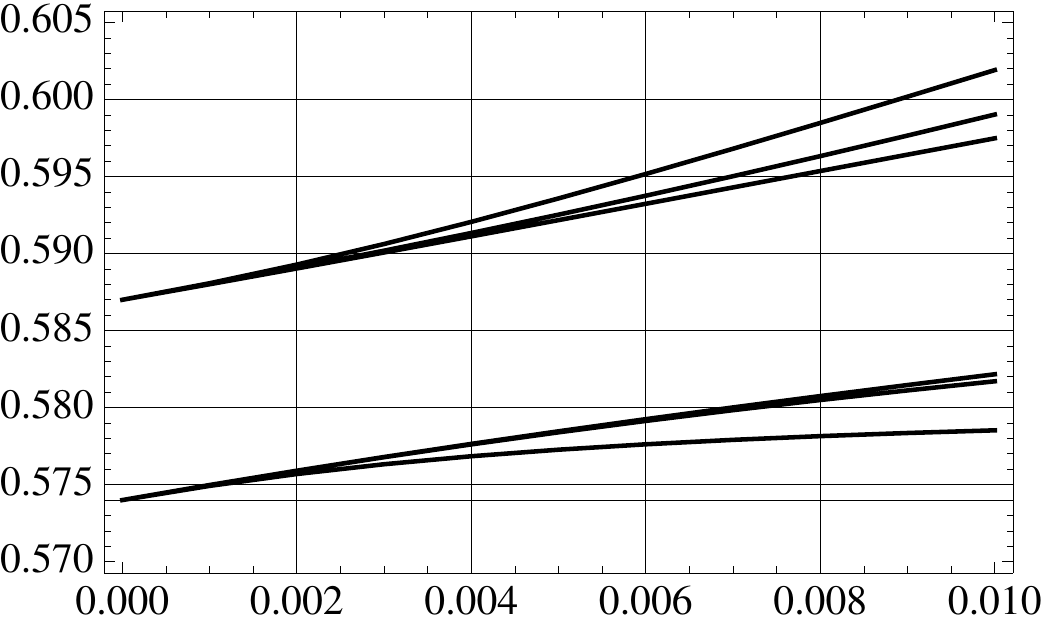}} \\
  & Beam-beam parameter, $\xi$
\end{tabular}
\caption{Coherent mode tunes vs. beam-beam parameter calculated with the
linearized model; $\qp = 0.587$, $\qa =0.574$, $\xi = \xip = \xia$.}
\label{fig:MatrixTunes}
\end{figure}

We used a simple matrix formalism to compute the eigenmode tunes of
the system of colliding bunches. Besides employing the rigid bunch
approximation, one more simplification was used. The complete
description of the system would require modeling the interaction of
72~bunches at 138 collision points. The analysis of such a system can
be quite complex. Observations and analytical estimates showed that
the difference in tunes between individual bunches was small compared
to the beam-beam tune shift. Thus, as a first approximation, it is
possible to neglect long-range interactions.  This reduces the system
to 6~bunches (3 in each beam) colliding at two head-on interaction
points. In the following discussion, we restrict betatron oscillations
to one degree of freedom. Because the system has 3-fold symmetry, the
1-turn map transporting the 12-vector of dipole moments and momenta of
the system of 6~bunches can be expressed as follows:
\begin{equation}
M = M_\mathrm{BB3} \  M_\mathrm{T3} \  M_\mathrm{BB2} \  M_\mathrm{T2}
\  M_\mathrm{BB1} \  M_\mathrm{T1} ,
\end{equation}
where $M_\mathrm{TN}$ ($N =1,2,3$) are the 2$\times$2 block-diagonal,
12$\times$12 matrices transporting phase space coordinates through the
accelerator arcs, and $M_\mathrm{BBN}$ are the matrices describing
thin beam-beam kicks at the IPs. Although there are only
2~interactions per bunch, 3~collision matrices are used to describe a
one-turn map of the system of 6~bunches. This construction represents
the time propagation of the bunch coordinates through one turn with
break points at the CDF (B0), D0 and F0 locations in the machine. If
on a given step the bunch is at B0 or D0, its momentum coordinate is
kicked according to the distance between the centroids of this bunch
and of the opposing bunch. If the bunch is at F0 (1/3 of the
circumference from B0 and D0), where the beams are separated, its
momentum is unchanged.  The eigentunes of the 1-turn map can then be
computed numerically. We will use the symbols \xip\ and \xia\ for the
beam-beam parameters of protons and antiprotons; $\beta$ is the
amplitude function at the IP. The Yokoya factor~\cite{Yokoya:PA:1990,
  Yokoya:PRSTAB:2000} is assumed to be equal to~1.

This model provides a quick estimate of the expected values of the
coherent beam-beam mode tunes for a given set of machine and beam
parameters. In Figure~\ref{fig:MatrixTunes}, an example of the
dependence of the 6~eigenfrequencies on the beam-beam parameter per IP
is presented. As one would expect, at small values of $\xi$ (uncoupled
oscillators) the mode frequencies approach the bare lattice tunes; in
this case, 0.587 for protons and 0.574 for antiprotons. When the total
beam-beam parameter exceeds the difference between the lattice tunes,
the modes are split and their symmetry approaches that of the
conventional $\sigma$ and $\pi$~modes.  The parameters of this
calculation are taken to resemble those of the beginning of the
Tevatron Store~7754, when the beam-beam parameter was $\xi = \xia =
\xip = 0.01$. A comparison with data is given in the results section
(Figure~\ref{fig:store7754_evolution}).

A more complete description of coherent oscillations was provided by
numerical simulations based on the code
\code{BeamBeam3D}~\cite{Qiang:PRSTAB:2002}. \code{BeamBeam3D} is a
fully parallelized 3-dimensional code allowing for self-consistent
field calculations of arbitrary distributions and tracking of multiple
bunches. Transport from one IP to the other is done through linear
transfer maps. The electromagnetic fields generated by the beams are
calculated from the Poisson equation using a shifted Green-function
method efficiently computed with a fast-Fourier-transform (FFT)
algorithm on a uniform grid.

The measured beam intensities and emittances were used in the
simulation. Lattice parameters were measured on the proton orbit. The
bare lattice tunes were derived from the main quadrupole currents. Due
to the asymmetry of the collision IPs in the Tevatron, the bunches
coupled by groups of~3 through the head-on interactions. In the
simulations, 3~bunches per beam were therefore tracked to reproduce
the spectrum of centroid oscillations. A comparison between the
calculated and measured spectra for the case of Tevatron Store~7754 is
discussed in the results section and shown in
Figure~\ref{fig:store7754_evolution}.

\section{APPARATUS}

\begin{figure}[t!]
\centering
\includegraphics[width=\columnwidth]{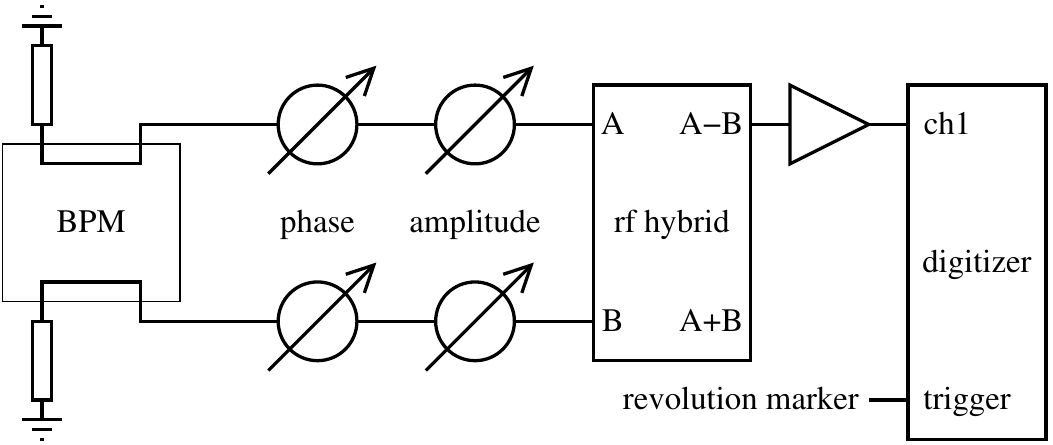}
\caption{Schematic diagram of the apparatus.}
\label{fig:apparatus}
\end{figure}

The system for the detection of transverse coherent modes
(Figure~\ref{fig:apparatus}) was based on the signal from a single
vertical beam-position monitor (BPM) located near the CDF interaction
point, in a region where the vertical amplitude function at collisions
was $\beta_y = \q{880}{m}$. The BPM was a stripline pickup, with two
plate outputs ($A$ and $B$) for each of the two counterpropagating
beams.

In the Tevatron, protons and antiprotons shared a common vacuum
pipe. Outside of the interaction regions, their orbits wrapped around
each other in a helical arrangement. Therefore, bunch centroids were
several millimeters away from the BPM's electrical axis. Typically,
the peak-to-peak amplitude of the proton signal was 10~V on one plate
and 5~V on the other, whereas the signal of interest was of the order
of a few millivolts.  For this reason, it was necessary to equalize
the $A$ and $B$ signals to take advantage of the full dynamic range of
the digitizer. Equalization also reduced false transverse signals due
to trigger jitter, as discussed below. The phase and attenuation of
each signal was manually adjusted by minimizing the $A-B$ output of
the rf hybrid circuit. If necessary, fine-tuning could be done by
displacing the beam with a small orbit bump.  Orbits at collisions
were stable over a time scale of weeks, and this manual adjustment did
not need to be repeated often.

The difference signal from the hybrid was amplified by 23~dB and sent
to the digitizer. We used a 1-channel, 1-V full range, 10-bit
digitizer with time-interleaved analog-to-digital converters
(ADCs). It sampled at 8~gigasamples/s (GS/s) and stored a maximum of
1024~megasamples (MS) or 125,000 segments. (Due to a firmware problem,
only half of the segments were used in the experiments described
below.) The 47.7-kHz Tevatron revolution marker was used as trigger,
so we refer to `segments' or `turns' interchangeably. Typically, we
sampled at 8~GS/s (sample period of 125~ps), i.e.\ 150~slices for each
19-ns rf bucket. At this sampling rate, one could record waveforms of
1~bunch for 62,500 turns, 2~bunches for 52,707 turns, or 12~bunches
for 12,382 turns, depending on the measurement of interest. A C++
program running on the front-end computer controlled the digitizer
settings, including its delay with respect to the Tevatron revolution
marker.

The recorded output data contained the raw ADC data together with the
trigger time stamps and the delay of the first sample with respect to
the trigger. Timing information had an accuracy of about 15~ps, and it
was extremely important for the synchronization of samples from
different turns.

To enhance the signal, the beam was excited with a few watts of
band-limited noise (`tickling') for about 1~s during the measurement.
The measurement cycle consisted of the following steps: digitizer
setup, tickler turn-on, acquisition start, tickler turn-off, and
acquisition stop. The cycle took a few seconds. The procedure was
parasitical and it did not adversely affect the circulating beams,
even at the beginning of regular collider stores, with luminosities
around \q{3.5\times 10^{32}}{events/(cm^2\, s)}. When repeating the
procedure several times, the Schottky monitors occasionally showed
some activity, but no beam loss was observed.

\section{DATA ANALYSIS}

\begin{figure*}[b!]
\centering
\includegraphics[width=0.98\textwidth]{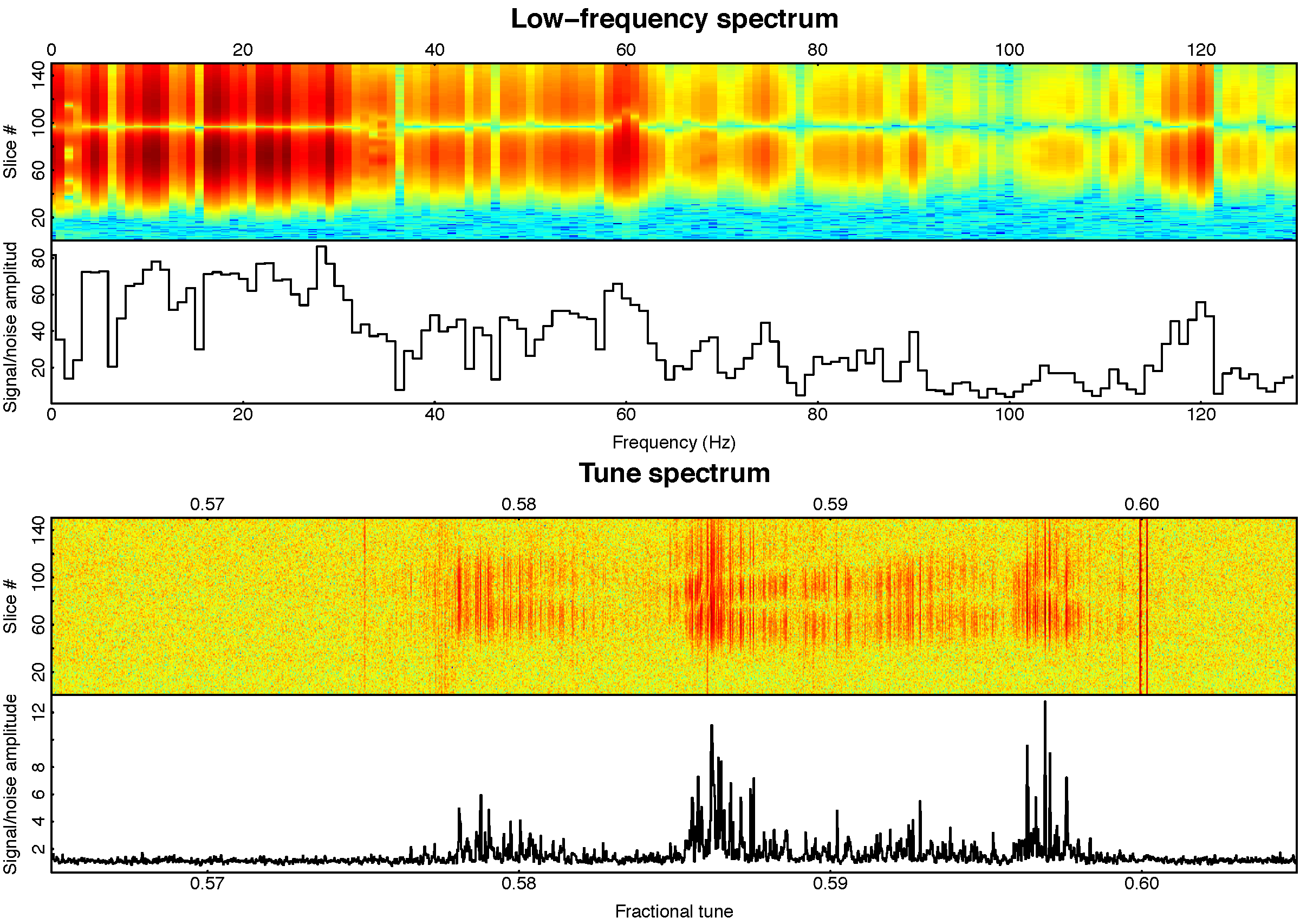}
\caption{Example of frequency spectra for antiprotons from data taken
  during Store~7754. Two selected regions of the spectrum are shown:
  below 130~Hz (top two plots) and around $(\q{47.7}{kHz}) \times
  (1-0.585) = \q{20}{kHz}$ (bottom two plots). The color plots
  represent the Fourier amplitude (in logarithmic scale) vs.\
  frequency for each of the 150 125-ps slices. The black traces are
  the average amplitudes of the signal slices divided by those of the
  background slices.}
\label{fig:analysis_example}
\end{figure*}

\begin{figure*}[p]
\centering
\includegraphics[width=\textwidth]{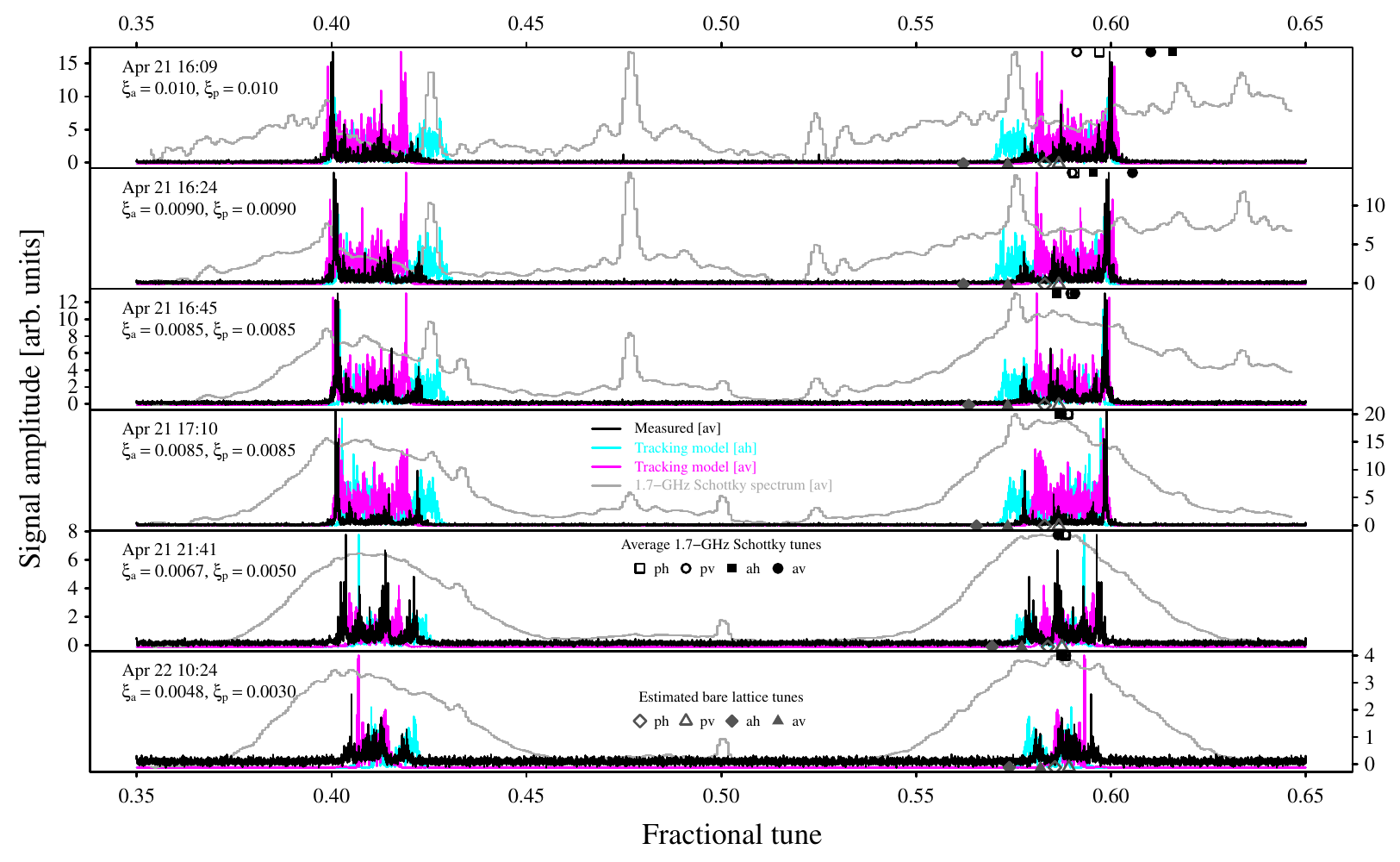}
\includegraphics[width=\textwidth]{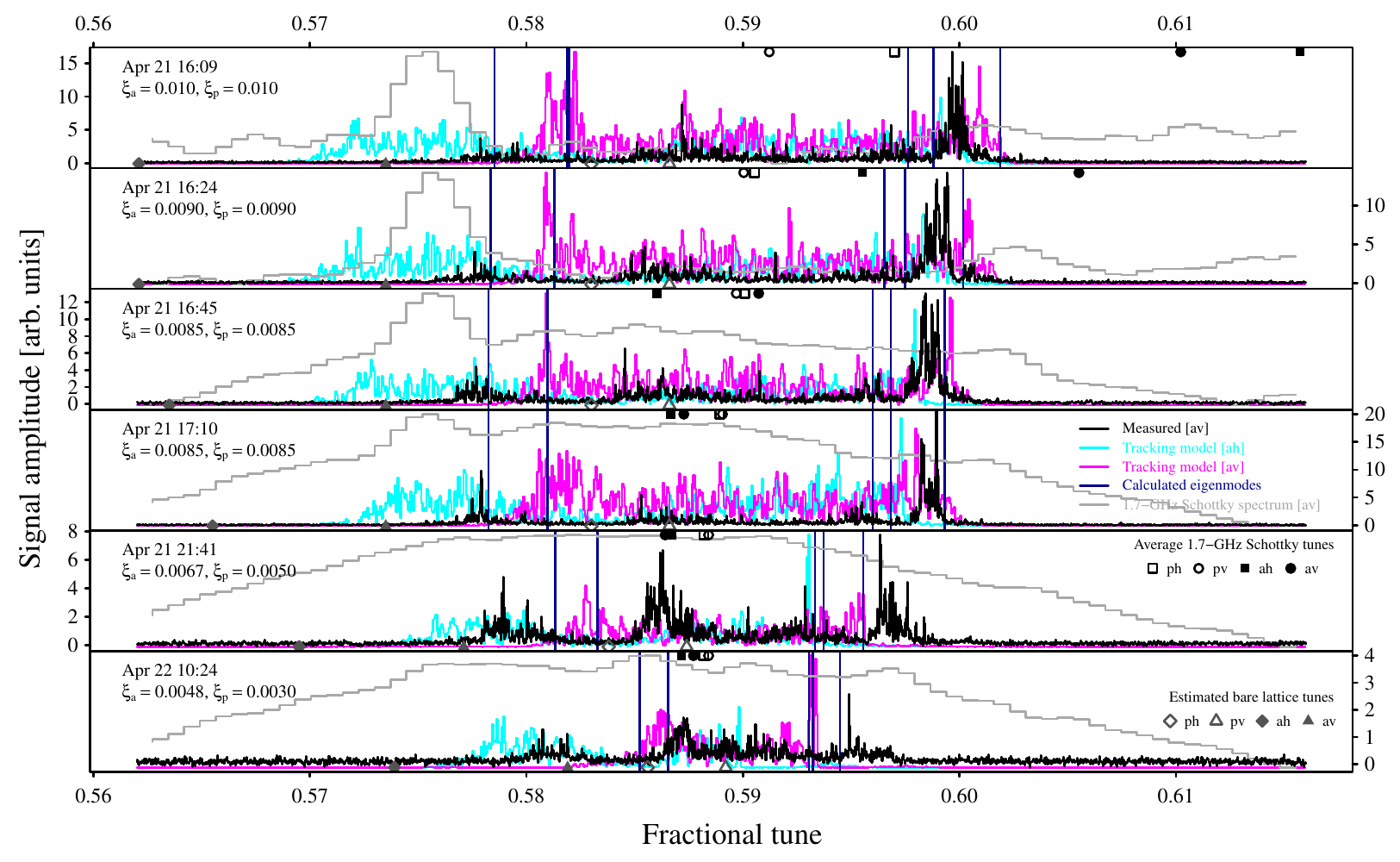}
\caption{Evolution of vertical coherent beam-beam modes for antiproton
  bunch~A13 during the course of Store~7754.}
\label{fig:store7754_evolution}
\end{figure*}

Data was analyzed offline using the multi-platform, open-source
\code{R}~statistical package~\cite{R:2010}. The distribution of
differences between trigger time stamps from consecutive turns was
used to obtain the average revolution frequency (47713.11~Hz at
980~GeV). From it, the nominal or `ideal' trigger time stamps for each
turn were calculated. The distribution of trigger offsets, i.e. the
differences between measured and nominal time stamps, is a measure of
the jitter in the revolution marker. The root mean square of the
distribution was usually less than 0.2~ns.  The delay between trigger
time and the time stamp of the first sample was also recorded with an
accuracy of 15~ps. The sum of trigger offset and first-sample delay is
the correction by which each sample in a segment is to be shifted in
time to be aligned with the other segments. For each turn and each
bunch, the signal was interpolated with a natural spline and shifted
in time according to this correction. One undesirable effect of this
synchronization algorithm is that a few slices (usually not more than
3) at each edge of the bucket become unusable, as they cannot be
replaced with real data. The synchronization of turns is extremely
important, as the jitter in trigger time translates into a false
transverse oscillation where the difference signal has a slope. If the
BPM plates are not perfectly balanced, jitter of even a fraction of a
nanosecond can raise the noise floor by several decibels and
compromise the measurement.

Bunch oscillations were dominated by low-frequency beam jitter
attributable to mechanical vibrations~\cite{Baklakov:PAC:1999,
  Shiltsev:JINST:2011}. The range of amplitudes was inferred from
comparisons with the regular Tevatron BPM system and corresponds to
about \q{\pm25}{\mu m}. This low-frequency jitter did not affect the
measurements of coherent beam-beam modes directly, but it reduced the
available dynamic range. A high-pass filter and more amplification may
be used to improve the system.

For each bunch, the signal of each individual slice vs.\ turn number
was Fourier transformed. Frequency resolution is determined by the
number of bins in the FFT vector and it is limited to 62,500 turns,
corresponding to $1.6\times 10^{-5}$ of the revolution frequency or
0.8~Hz. The data was multiplied by a Slepian window of rank~2 to
confine leakage to adjacent frequency bins and suppress it below
$10^{-5}$ in farther bins~\cite{Press:NR:2007}. When the full
frequency resolution was not needed, the FFT vectors were overlapped
by about 1/3 of their length to reduce data loss from windowing, and
the resulting spectral amplitudes were averaged. Calculations took
about 20~s per bunch for 62,500~turns and 150~slices per bunch on a
standard laptop computer. Processing time was dominated by the
synchronization algorithm.

The noise level was estimated by observing the spectra without
beam. The spectra showed a few sharp lines in all slices. These lines
were attributed to gain and offset differences between the
time-interleaved ADCs themselves and to timing skew of their
clocks. To improve the signal-to-noise ratio, and to suppress
backgrounds unrelated to the beam such as the spurious lines from the
digitizer, a set of signal slices (near the signal peaks) and a set of
background slices (before the arrival of the bunch) were
defined. Amplitude spectra were computed for both signal and
background slice sets, and their ratio was calculated. The ratios are
very clean, with some additional variance at the frequencies
corresponding to the narrow noise spikes. Results are presented in
terms of these signal-to-background amplitude ratios.

Figure~\ref{fig:analysis_example} shows an example of analyzed
antiproton data, in two regions of the frequency spectrum: a
low-frequency region with the horizontal axis expressed in hertz (top
two plots) and a high-frequency region, in terms of the revolution
frequency or fractional tune. The 2-dimensional color plots show the
amplitude distribution for each of the 150 125-ps slices in
logarithmic scale. In this example, the signal slices are numbers
41--95 and 99--147. They are defined as the ones for which the
amplitude is above 10\% of the range of amplitudes. Background slices
are numbers 3--31 (amplitude below 2\% of range). The black-and-white
1-dimensional plots show the ratio between signal and background
amplitudes.  In the top plots of Figure~\ref{fig:analysis_example},
one can appreciate the strength of the low-frequency components. The
60-Hz power-line noise and its harmonics are also visible. The lines
around 34~Hz and 68~Hz are due to synchrotron oscillations leaking
into the transverse spectrum. The bottom plots of
Figure~\ref{fig:analysis_example} show the spectra of transverse
coherent oscillations. The vertical lines present in all slices in the
2-dimensinal plot, attributed to digitizer noise, are eliminated by
taking the ratio between signal and background slices. One can also
notice the small variance of the noise level compared to the amplitude
of the signal peaks.

In the 2-dimensional plots of Figure~\ref{fig:analysis_example}, one
may notice patterns in the oscillation amplitude as a function of
position along the bunch. These may be an artifact of the imperfect
synchronization between the~$A$ and $B$ signals, but they may also be
related to the physical nature of the coherent modes (i.e., rigid vs.\
soft bunch, head-on vs.\ long range). The phase of the oscillations as
a function of frequency and bunch number may also provide physical
insight.

\section{RESULTS}
\label{sec:results}

Transverse coherent mode spectra were measured for both proton and
antiproton bunches under a wide range of experimental
conditions~\cite{Stancari:PRSTAB:2012}. In this section, we focus on
the observation of coherent beam-beam modes, on their evolution over the
course of a collider store, and on comparisons with analytical and
numerical models.

An illustration of the evolution of transverse coherent modes during a
collider store is shown in Figure~\ref{fig:store7754_evolution} for
vertical antiproton oscillations. The top plot covers a wide range of
fractional tunes, while the bottom one shows the details near the
betatron frequencies. The black line represents the measured
spectra. The gray histogram shows the measured 1.7-GHz antiproton
vertical Schottky spectra for comparison. The cyan and magenta lines
are the antiproton horizontal (ah) and antiproton vertical (av)
spectra calculated with the \code{BeamBeam3D} code. The bottom plot
shows the~6 calculated rigid-bunch modes as vertical dark blue
lines. Markers are used to indicate the average Schottky tunes (black)
and the estimated bare lattice tunes (dark gray) for protons and
antiprotons, both horizontally and vertically (ph, pv, ah, and
av). The first 4~spectra were acquired within about 1~hour after the
beams were brought in collision. The 5th plot was taken after about
6~hours, whereas the last plot was taken at the end of the store, just
before the beams were dumped. The calculated beam-beam parameters per
IP, \xia\ and \xip, are printed on the left side of each plot.

Over the course of a store, the lattice tunes need to be periodically
adjusted to keep the average incoherent tune close to the desired
working point.  Except for the last two measurements, which may be
affected by the evolving linear coupling and by a slight
miscalibration of the tune settings, the estimated lattice tune (dark
gray triangles) lies below the first group of eigenmodes, as expected.

One can clearly see how, as the beam-beam force weakens, the spread in
coherent modes decreases, and so does the amplitude of the $\pi$~mode
(near 0.60). The asymmetries between the beams, the large number of
bunches, and the multiple collision points give rise to a rich
spectrum of oscillations.

A comparison with the Schottky spectra reveals many common coherent
spikes. The ones at 0.475/0.525, visible in both the Schottky spectrum
and in the digitized-BPM spectrum, are unexplained. Because of the
distortions of the Schottky spectrum at the beginning of the store,
the present system provides a better indication of the tune
distribution under these conditions.

The predicted eigenfrequencies of the simplified rigid-bunch model are
close to the measured peaks.  Obviously, the measured spectra are
richer than those predicted by the simplified model, and a complete
explanation requires a more detailed description of the beam dynamics,
such as the one based on the 3-dimensional strong-strong
code.  The results of the \code{BeamBeam3D}
simulations are very similar to the data. The comparison between data
(vertical) and simulations (both horizontal and vertical) suggests
that the effect of coupling, not included in simulations, is
non-negligible and may account for some of the discrepancies.

\section{CONCLUSIONS}

A system was developed to measure the spectra of coherent beam-beam
oscillations of individual bunches in the Fermilab Tevatron
collider. It is based on the analysis of the digitized signal from a
single beam-position monitor. It requires applying band-limited noise
to the beam, but an extension of its dynamic range is possible, if
needed, so as to operate without excitation.

The device has a response time of a few seconds, a frequency
resolution of $1.6\times 10^{-5}$ in fractional tune, and it is
sensitive to oscillation amplitudes of 60~nm. In terms of sensitivity,
resolution, and background level, it provides a very clean measurement
of coherent oscillations in hadron machines. The system is
complementary to Schottky detectors and transfer-function measurements
as a diagnostic tool for tunes, tune spreads, and beam-beam effects.
At the beginning of a collider store, when strong coherent lines
distort the incoherent Schottky tune spectrum, the present system may
provide a more accurate indication of betatron tunes.

Coherent oscillations in the Tevatron were stable, probably thanks to
the different intensities of the two beams, their tune separation, and
chromaticity. The average amplitude of the oscillations around the
ring was estimated to be of the order of 20~nm. Patterns in the
oscillation amplitude as a function of position along the bunch were
observed. They may be related to the physical nature of the coherent
modes. The phase of the oscillations as a function of frequency and
bunch number may also provide physical insight, but it was not
considered in this analysis.

A simplified collision model was used to calculate normal mode
frequencies and to show their dependence on beam-beam coupling. Some
scenarios were simulated using the self-consistent 3-dimensional
strong-strong numerical code \code{BeamBeam3D}. Models were compared
with observations made over the course of a collider store, as the
strength of the beam-beam force decreased with time. Some
discrepancies were observed, but the overall agreement was
satisfactory considering the uncertainties on the antiproton
parameters, such as lattice tunes and coupling, and their variation
over time.

\section{ACKNOWLEDGMENTS}

The authors would like to thank V.~Kamerdzhiev (Forschungszentrum
J\"ulich, Germany), F.~Emanov (Budker Institute for Nuclear Physics,
Novosibirsk, Russia), Y.~Alexahin, B.~Fellenz, V.~Lebedev, G.~Saewert,
V.~Scarpine, A.~Semenov, and V.~Shiltsev (Fermilab) for their help and
insights.

Fermi Research Alliance, LLC operates Fermilab under Contract
No.~DE-AC02-07-CH-11359 with the United States Department of Energy
(US DOE). Brookhaven National Laboratory is operated by Brookhaven
Science Associates, LLC under Contract No.~DE-AC02-98-CH-10886 with
the US DOE. This work was partially supported by the US LHC
Accelerator Research Program (LARP).

\end{document}